\documentclass[preprint,amsmath,amssymb,11pt,english,aps,showpacs]{revtex4}
\usepackage{graphicx}
\usepackage{graphics}
\usepackage{amsmath}
\usepackage{dcolumn}
\usepackage{amssymb}
\usepackage{bm}
\usepackage[latin1]{inputenc}

\begin{document}
\title{Relativistic Einstein-Podolsky-Rosen Correlations in curved spacetime via Fermi-Walker Transport}
\author{Knut Bakke$^{1}$, Alexandre M. de M. Carvalho$^{2}$ and Claudio Furtado$^{1}$}
\email{kbakke@fisica.ufpb.br,furtado@fisica.ufpb.br} 
\affiliation{$^{1}$Departamento de F\'{\i}sica, Universidade Federal da Para\'{\i}ba, Caixa Postal 5008, 58051-970, Jo\~ao Pessoa, Pb, Brazil\\
$^{2}$Departamento de F\'{\i}sica, Universidade Estadual de Feira de 
Santana, 
BR116-Norte, Km 3, 
44031-460, Feira de Santana, BA, Brazil}

\begin{abstract}
We present a geometric description to study the relativistic EPR correlations in curved spacetime background given by the application of the Fermi-Walker transport in the relativistic EPR states and we  show that its result has the same effect as  the applications of successive infinitesimal Lorentz boosts in the relativistic EPR states. We  also show that the expression for the Bell inequality due to the Fermi-Walker transport is equivalent to the expression demonstrated by Terashima and Ueda \cite{TU2}, where the degree of violation of the Bell inequality is dependent of the angle of the Wigner rotation. This geometrical approach to study the relativistic EPR correlations is a promissing formulation to investigate the EPR correlations in the general relativity background.   
\end{abstract}
\maketitle

\section{Introduction}
$ $ 



One of the most famous discussions about the interpretation of the quantum theory was done by Einstein, Podolsky and Rosen \cite{epr}, Bohr \cite{epr2}, Bohm \cite{epr3} and Bell \cite{epr4} and became known as the Einstein-Podolsky-Rosen (EPR) paradox. At present days, quantum mechanical properties like   like  entanglement and  non-locality have been a fundamental feature in the development of  quantum cryptography \cite{epr5,epr6} and quantum teleportation \cite{epr7}. 

In the framework of the relativistic quantum theory, the EPR correlations have been discussed both in special relativity \cite{2,3,4,5,rl3,6,7,rl1,1,rl2} and  in general relativity \cite{mensky,TU2,tu3,tu4,tu5,bf} context. One of the most importants effects is the precession of the spins of the relativistic EPR particles in relation to their initial configuration in the local reference frame of the observers due to Lorentz transformations. This precession is known as Wigner rotation \cite{wig,Weinberg2} and configures an apparent deterioration of the initial correlations between the spins. An important consequence of the spin precession in relativistic EPR correlation is the violation of the Bell's inequality. In \cite{2,4,5,rl3,rl2} it was shown that the violation of the Bell's inequality decreases due the relativistic motion of the particle and in \cite{TU2,tu3,bf}, the decrease of the Bell inequality is made by the gravitational field, the relativistic motion of the particle and the position of the observers. However, with appropriate choices of the direction of the spin axis of measurements in the local reference of the observers, the perfect spin correlations and the maximal violation of the Bell's inequality can be recuperated. 

In special relativity, some works \cite{Weinberg2,2,3,4,5,6,7,rl3} have been considered the relativistic EPR states as given with the information of the momentum of the EPR particles and the spin of these particles, where the degrees of freedom of the momentum are correlated with the degree of freedom of the spins. This discussion of the basis of the EPR states in special relativity was done clearly in  reference \cite{7}. However, in the curved spacetime background, the definition of the particle states are not unique as pointed out in \cite{bd}. To solve this problem, we need to build a local reference frame for observers where the spacetime is locally identical to the Minkowisky spacetime. In that way, the states of the particle are locally well defined and the spinors transform under local Lorentz transformations. In general relativity, the relativistic EPR states considered in \cite{TU2,tu3} had taking into account the position of the observers. The expressions for these states are
\begin{eqnarray}
\left|\psi^{\pm}\right\rangle&=&\frac{1}{\sqrt{2}}\left\{\left|p^{a}_{+}\left(x\right),\uparrow;x\right\rangle\left|p^{a}_{-}\left(x\right),\downarrow;x\right\rangle\pm\left|p^{a}_{+}\left(x\right),\downarrow;x\right\rangle\left|p^{a}_{-}\left(x\right),\uparrow;x\right\rangle\right\},\\
\left|\phi^{\pm}\right\rangle&=&\frac{1}{\sqrt{2}}\left\{\left|p^{a}_{+}\left(x\right),\uparrow;x\right\rangle\left|p^{a}_{-}\left(x\right),\uparrow;x\right\rangle\pm\left|p^{a}_{+}\left(x\right),\downarrow;x\right\rangle\left|p^{a}_{-}\left(x\right),\downarrow;x\right\rangle\right\},
\label{1}
\end{eqnarray}  
where $p^{a}\left(x\right)$ is defined in the local reference frame of the observers, $\sigma=\uparrow,\downarrow$ indicates the spins of the particles and $x$ denotes the position of the observers. By definition, each state above transforms under local Lorentz transformations in the spin-half representation at each local reference frame of the observers. The application of successive infinitesimal local Lorentz transformations at each point of the curved spacetime makes  the one-particles states given in (\ref{1}) transform as \cite{TU2}
\begin{eqnarray}
U\left(\Lambda\left(x\right)\right)\left|p^{a}\left(x\right),\sigma;x\right\rangle=\sum_{\sigma'}\,D^{1/2}_{\sigma\sigma'}\left(W\left(x\right)\right)\left|\Lambda p^{a}\left(x\right),\sigma';x\right\rangle,
\label{1.1}
\end{eqnarray} 
with $\Lambda^{a}_{\,\,\,b}\left(x\right)$ being the local Lorentz transformation and $W\left(x\right)$ the local Wigner rotation. The special characteristic of the Wigner rotation is that the momentum of the particles remain unchanged in the rest frame of the particles, but the spins of the particles undergo a precession due the accelerated motion of the particle and the change of the local reference frame at different points of  spacetime.

The first geometrical approach to study the relativistic EPR correlations in the presence of the gravitational field was proposed by Borzeskowski and Mensky \cite{mensky}. They showed that it may exist an approximate correlation and the correlated directions are connected with each other by the parallel transport along the world lines of the particles. However, Terashima and Ueda \cite{TU2,tu3} showed that, through the application of successive infinitesimal Lorentz transformations, if we take account of the gravity and the accelerated motion of the particles, the parallel transport does not give the perfect direction of the relativistic EPR correlations. 

In the last decade, the application of the Fermi-Walker transport in the Dirac equation was done in \cite{ryder}, where the spacetime has a torsion field. The components of the Fermi-Walker transport was incorporated in the Dirac equation via tetrad field, where the acceleration of the frames was given as a metric fluctuation. Anandan \cite{anandan2,anandan} introduced the Fermi-Walker transport in the context of the relativistic quantum mechanics and  general relativity through the WKB approximation adding a new term in the gravitational phase factor. He studied the interference effects of neutral particles in  curved spacetime with torsion without treating the acceleration of the observers as a fluctuation of the metric of  spacetime and showed that there is a quantum flux due the accelerated motion of the observers. 
With this in  mind, we investigate in this work, the relativistic EPR correlations in the context on general relativity using the Fermi-Walker transport introduced by Anandan in \cite{anandan2,anandan}.

In this paper, we  apply the Fermi-Walker transport defined in \cite{anandan2,anandan} on the EPR states given by expressions (\ref{1}). We also consider that the wave packet of each particle can be expanded as a locally plane wave superposition which satisfies the WKB approximation. The application of the Fermi-Walker transport  is used to carry the wave packet to different points of the curved spacetime where the local reference frame of the observers changes. The final result of the Fermi-Walker transport is identical to the application of successive infinitesimal Lorentz transformations, where the local Wigner rotation is obtained. We will see that the Fermi-Walker transport leaves the momenta of the particles unchanged in their rest frame, where the observers are placed as a result, we get that the phase acquired by the spatial part of the wave function of the particles, are precessioned in relation to the local reference frame, is identical to the angle of the Wigner rotation.   

We structure this paper as follow: in section II, we present the Fermi-Walker transport as introduced in \cite{anandan2,anandan}. In section III, we present the cosmic string background where the relativistic EPR particles will describe a circular motion. In section IV, we will study the behaviour of the relativistic EPR correlations in the cosmic string spacetime considering the WKB approximation. In section V, we discuss the Bell inequality in this geometrical approach. In section VI, we present our conclusions.

\section{the fermi-walker transport}
$ $ 
In general relativity, we need to define the spinor in curved spacetime introducing a local reference frame at each point \cite{TU2,bd}. Thus, in the local frame of the observers, the spinors transform under local Lorentz transformations. We can build the local reference frame of the observers defining a non-coordinate basis $\hat{\theta}^{a}=e^{a}_{\,\,\,\mu}\left(x\right)\,dx^{\mu}$, where its elements $e^{a}_{\,\,\,\mu}\left(x\right)$ satisfy the relation \cite{naka, bd}
\begin{eqnarray}
g_{\mu\nu}\left(x\right)=e^{a}_{\,\,\,\mu}\left(x\right)\,e^{b}_{\,\,\,\nu}\left(x\right)\,\eta_{ab},
\label{1.1}
\end{eqnarray}
where $\eta_{ab}=diag(-\,+\,+\,+\,)$ is the Minkowisky tensor. These elements of the non-coordinate basis are called \textit{tetrads or vierbein}. The greek indices indicate the spacetime indeces and the latin indexes indicate the local reference frame which runs $a,b,c=0,1,2,3$. The tetrads has a inverse defined as $dx^{\mu}=e^{\mu}_{\,\,\,a}\left(x\right)\,\hat{\theta}^{a}$, where 
\begin{eqnarray}
e^{a}_{\,\,\,\mu}\left(x\right)\,e^{\mu}_{\,\,\,b}\left(x\right)=\delta^{a}_{\,\,\,b};\,\,\,\,\,\,\,e^{\mu}_{\,\,\,a}\left(x\right)\,e^{a}_{\,\,\,\nu}\left(x\right)=\delta^{\mu}_{\,\,\,\nu}.
\label{1.2}
\end{eqnarray}

Hence, when we transport a vector at a point $x$ to a new point $x'$ and consider the action of external forces on the tetrads, but with the axis of the tetrads undergoing no torques during the transport, the law of transport on the tetrads will be the Fermi-Walker transport \cite{misner,synge,stph}
\begin{eqnarray}
\frac{D}{D\tau}\,e^{a}_{\,\,\,\mu}\left(x\right)=\frac{-1}{c^{2}}\left[a_{\mu}\left(x\right)\,U^{\beta}\left(x\right)-U_{\mu}\left(x\right)\,a^{\beta}\left(x\right)\right]\,e^{a}_{\,\,\,\beta}\left(x\right),
\label{1.3}
\end{eqnarray}   
where $\frac{D}{D\tau}$ is the covariant derivative. As discussed in \cite{TU2,tu3}, the evolution of each wave packet in  curved spacetime is determined by the application of successive infinitesimal Lorentz boosts at each point of spacetime making an open or closed path. When the Fermi-Walker transport is applied in the spinors, considering the WKB approximation \cite{anandan}, the wave function acquires a phase shift that can be described by the operator
\begin{eqnarray}
\Xi_{\lambda}=\hat{P}\exp\left(-i\int_{\lambda}e^{\,\,\,a}_{\mu}\,\,P_{a}+\frac{1}{2}\Omega_{\mu ab}\,\Sigma^{ab}\,dx^{\mu}\right),
\label{2}
\end{eqnarray}
where $P_{a}$ and $\Sigma^{ab}=\frac{i}{2}\left[\gamma^{a},\gamma^{b}\right]$ are the translation generator of the Poincar\'e group that acts on Hilbert space and the $\gamma^{a}$ matrices are  the flat spacetime Dirac matrices. This expression give us the notion of Fermi-Walker transport which was introduced by Anandan in \cite{anandan}. In this work we shall consider that there is no torsion in  spacetime, thus, our system is described by the Lorentz group. In this way, the Fermi-Walker transport is given only by the second term in the expression (\ref{2}) which we can rewrite as
\begin{eqnarray}
\Xi_{\lambda}=\hat{P}\exp\left(\frac{-i}{2}\int\Omega_{\mu ab}\,\Sigma^{ab}\,U^{\mu}\,d\tau\right),
\label{3}
\end{eqnarray} 
where $U^{\mu}=\frac{dx^{\mu}}{d\tau}$ is the four-velocity and $\Omega_{\mu\,\,\,b}^{\,\,\,a}$ is given by
\begin{eqnarray}
\Omega_{\mu\,\,\,b}^{\,\,\,a}=\omega_{\mu\,\,\,b}^{\,\,\,a}+\tau_{\mu\,\,\,b}^{\,\,a}.
\label{4}
\end{eqnarray}
The term $\omega_{\mu\,\,\,b}^{\,\,a}$ is the spin connection and its expression is \cite{naka,TU2,bd}
\begin{eqnarray}
\omega_{\mu\,\,\,b}^{\,\,a}=-e^{a}_{\,\,\,\nu}\left(x\right)\,\nabla_{\mu}\,e^{\nu}_{\,\,\,b}\left(x\right),
\label{5}
\end{eqnarray}
with $\nabla_{\mu}$ are the components of the covariant derivative of a vector. The term $\tau_{\mu\,\,\,b}^{\,\,a}$ was introduced in \cite{anandan2,anandan}, as (\ref{1.3}): 
\begin{eqnarray}
\tau_{\mu\,\,\,b}^{\,\,a}\left(x\right)=\frac{a^{\nu}}{c^{2}}\left[e_{\,\,\,\nu}^{a}\left(x\right)\,e_{b\mu}\left(x\right)-e_{\,\,\,\mu}^{a}\left(x\right)\,e_{b\nu}\left(x\right)\right].
\label{6}
\end{eqnarray}
and refers to the application of the Fermi-Walker transport in the local reference frame of the particles. The expression (\ref{6}) makes possible to investigate the influence of the accelerated motion of the particles in the phase shift of the wave function. We  show that the action of the Fermi-Walker transport on relativistic EPR states is equivalent to the application of infinitesimal Lorentz boosts on the relativistic EPR states at each point of the spacetime, \textit{i.e.}, the Fermi-Walker transport makes a spin precession identical to the Wigner rotation discussed in \cite{TU2,bf}.

\section{cosmic string background}
$ $ 

In this section, we  consider that the EPR pair of particles are moving in the cosmic string spacetime. We take the simplest form of the cosmic string given by a straight line as done in \cite{bf}. The line element of cosmic string is given by
\begin{equation}
ds^{2}=-c^{2}dt^{2}+d\rho^{2}+dz^{2}+\alpha^{2}\rho^{2}d\varphi^{2},  
\label{7}
\end{equation}
where $\alpha=1-4G\nu$ is the angular deficit factor defined in the range $0<\alpha<1$, with $\nu$ being the linear mass density. The azimuthal angle varies in the interval: $0\leq\varphi<2\pi$. The deficit angle can assume only values for which $\alpha<1$ (unlike of this, in \cite{kat,furt}, it can assume values greater than 1, which correspond to an anti-conical space-time with negative curvature). This geometry possess a conical singularity represented by the following curvature tensor
\begin{eqnarray}
\label{curv}\label{curva}
R_{\rho,\varphi}^{\rho,\varphi}=\frac{1-\alpha}{4\alpha}\delta_{2}(\vec{r}),
\end{eqnarray}
where $\delta_{2}(\vec{r})$ is the two-dimensional delta function. This behavior of the curvature tensor is denominated conical singularity \cite{staro}. The conical singularity gives rise to the curvature concentrated on the cosmic string axis. In all  other places the curvature is null.

Our choice of the local reference frame which is at rest and its spatial axis do not rotate is given by the tetrads 
\begin{eqnarray}
e^{0}_{\,\,\,t}\left(x\right)=1;\,\,\,\,e^{1}_{\,\,\,\rho}\left(x\right)=1;\,\,\,\,e^{2}_{\,\,\,z}\left(x\right)=1;\,\,\,\,e^{3}_{\,\,\,\varphi}\left(x\right)=\alpha\rho.
\label{7.1}
\end{eqnarray} 

At this moment, we consider that the particles are moving in circular motion with a constant ratio $\rho$ around the symmetry axis of the cosmic string. The proper time of each particle is given by \cite{bf}
\begin{eqnarray}
\tau=\frac{\alpha\rho}{c}\frac{\Phi}{\sinh\xi}.
\label{11}
\end{eqnarray}

The four-velocity of each particle is given by the components \cite{bf}
\begin{equation}
U^{t}\left(x\right)=c\cosh\xi,\,\,\,\,\,\,\,\, U^{\varphi}\left(x\right)=\frac{c}{\alpha\rho}\sinh\xi 
\label{8}
\end{equation}
with $\frac{v}{c}=\tanh\xi$ and the acceleration of the particles has one non-null component \cite{bf} 
\begin{eqnarray}
a^{\rho}\left(x\right)=\frac{-c^{2}}{\rho}\sinh^{2}\xi .
\label{9}
\end{eqnarray}
The $1$-form connections $\tau_{\mu\,\,\,b}^{\,\,a}\left(x\right)$ of  expression (\ref{6}) have the following non-null components
\begin{eqnarray}
\tau_{t\,\,\,1}^{\,\,0}=\tau_{t\,\,\,0}^{\,\,1}=\frac{-a^{\rho}}{c^{2}},\,\,\,\,\,\tau_{z\,\,\,2}^{\,\,1}=-\tau_{z\,\,\,1}^{\,\,2}=\frac{a^{\rho}}{c^{2}},\,\,\,\,\,\,\tau_{\varphi\,\,\,3}^{\,\,1}=-\tau_{\varphi\,\,\,1}^{\,\,3}=\alpha\rho\,\frac{a^{\rho}}{c^{2}},
\label{9.1}
\end{eqnarray}
and the spin connections (\ref{5}) has two non-null components 
\begin{eqnarray}
\omega_{\varphi\,\,\,1}^{\,\,3}=-\omega_{\varphi\,\,\,3}^{\,\,1}=\alpha.
\label{9.2}
\end{eqnarray}
In that way, we have that the non-null components of the object $\Omega_{\mu\,\,\,b}^{\,\,\,a}\left(x\right)$, given in (\ref{4}), are 
\begin{eqnarray}
\Omega_{t\,\,\,1}^{\,\,0}=\frac{-a^{\rho}}{c^{2}};\,\,\,\,\,\Omega_{z\,\,\,2}^{\,\,1}=\frac{a^{\rho}}{c^{2}};\,\,\,\,\,\Omega_{\varphi\,\,\,3}^{\,\,1}=-\alpha\left(1-\frac{\rho\,a^{\rho}}{c^{2}}\right).
\label{10}
\end{eqnarray}

Now we have all information about the topology of the spacetime and the motion of the particles around the symmetry axis of the cosmic string. With this in our hands we will study in next section the behaviour of the EPR correlation in this system.

\section{epr correlations}
$ $ 

In this section, we consider that there are two observers and an EPR source on the $z=cont$ plane, where their positions are given by the azimuthal angles $\varphi=\pm\Phi$ and $\varphi=0$, respectively.  So, we consider that an EPR pair of particles are emitted from the source in opposite directions in circular motion, whose initial states are described for one of the EPR states (\ref{1}):
\begin{eqnarray}
\left|\psi_{-}\right\rangle&=&\frac{1}{\sqrt{2}}\left\{\left|\vec{p}_{+}\left(0\right),\uparrow;\varphi=0\right\rangle\left|\vec{p}_{-}\left(0\right),\downarrow;\varphi=0\right\rangle-\left|\vec{p}_{+}\left(0\right),\downarrow;\varphi=0\right\rangle\left|\vec{p}_{-}\left(0\right),\uparrow;\varphi=0\right\rangle\right\},
\label{4.1}
\end{eqnarray}
where the four-momenta of each particle in the local reference frame are given as in \cite{bf}
\begin{eqnarray}
p^{a}_{\pm}\left(0\right)=\left(mc\,\cosh\xi,0,0,\pm mc\,\sinh\xi\right)
\label{4.2}
\end{eqnarray}

After the emission of and EPR pair of particles, we apply the Fermi-Walker transport (\ref{3}) in the state (\ref{4.1}). We consider the WKB approximation when we transport the state (\ref{4.1}) from the point $\varphi=0$ in the source to $\varphi=\pm\Phi$ where the observers are placed. Considering that the observers are in the rest frame of the particles (\ref{7.1}), the application of the Fermi-Walker transport is analogue to the application of infinitesimal Lorentz boots at each point of the spacetime, \textit{i.e.}, the four-momentum of the particles (\ref{4.2}) remain unchanged, but the spins of the particles undergo a precession in the local reference frame of the observers in relation to their initial configuration. In this way, we apply the Fermi-Walker transport at each spin subspace of the EPR state (\ref{4.1}). The $\gamma$ matrices become the Pauli matrices at each spin subspace and the expression for the Fermi-Walker transport becomes
\begin{eqnarray}
\Xi&=&\exp\left(\frac{\eta_{1}}{2}\,\sigma^{1}+\frac{\eta_{2}}{2}\,i\sigma^{2}\right)=\exp\left(\frac{\Gamma}{2}\right),
\label{4.3}
\end{eqnarray}
where we define the parameters
\begin{eqnarray}
\eta_{1}&=&-\alpha\,\Phi\,\sinh\xi\,\cosh\xi;\,\,\,\,\eta_{2}=-\alpha\,\Phi\,\cosh^{2}\xi.
\end{eqnarray}
Expression (\ref{4.3}) becomes
\begin{eqnarray}
\Xi=\cosh\left(\frac{\gamma}{2}\right)+\frac{\Gamma}{\gamma}\,\sinh\left(\frac{\gamma}{2}\right)
\label{4.4}
\end{eqnarray}
with the $\Gamma$ matrix being 
\begin{eqnarray}
\Gamma=\left(
\begin{array}{cc}
0 & \eta_{1}+\eta_{2}\\
\eta_{1}-\eta_{2} & 0\\
\end{array}\right),
\label{4.4.1}
\end{eqnarray}
where the parameter $\gamma$ in (\ref{4.4}) is given by   
\begin{eqnarray}
\gamma=\sqrt{\eta_{1}^{2}-\eta_{2}^{2}}=\pm\,i\,\alpha\,\Phi\,\cosh\xi.
\label{4.5}
\end{eqnarray}

Hence, applying the operator (\ref{4.3}) in the spin states of each particle, we obtain
\begin{eqnarray}
\Xi\,\left|\uparrow\right\rangle=\cosh\left(\frac{\eta}{2}\right)\,\left|\uparrow\right\rangle\pm\frac{\eta_{1}-\eta_{2}}{\gamma}\,\sinh\left(\frac{\eta}{2}\right)\,\left|\downarrow\right\rangle,
\label{4.6}
\end{eqnarray}
and
\begin{eqnarray}
\Xi\,\left|\downarrow\right\rangle=\cosh\left(\frac{\eta}{2}\right)\,\left|\downarrow\right\rangle\pm\frac{\eta_{1}+\eta_{2}}{\gamma}\,\sinh\left(\frac{\eta}{2}\right)\,\left|\uparrow\right\rangle,
\label{4.7}
\end{eqnarray}
where the up sign indicates the first particle and the down sign the second particle. Appling the Fermi-Walker transport via expressions (\ref{4.6}) and (\ref{4.7}) in the initial state (\ref{4.1}), we  obtain that the final state in the local reference frame of the observers
\begin{eqnarray}
\left|\Psi\right\rangle&=&\cosh\gamma\,\left|\psi^{-}\right\rangle-\frac{\eta_{1}}{\gamma}\,\sinh\gamma\,\left|\phi^{-}\right\rangle-\frac{\eta_{2}}{\gamma}\,\sinh\gamma\,\left|\phi^{+}\right\rangle\nonumber\\
&=&\cos\theta\,\left|\psi^{-}\right\rangle+e^{\mp i\pi/2}\,\sin\theta\left\{\sinh\xi\,\left|\phi^{-}\right\rangle+\cosh\xi\,\left|\phi^{+}\right\rangle\right\},
\label{4.8}
\end{eqnarray}   
where the states $\left|\psi^{\pm}\right\rangle$ and $\left|\phi^{\pm}\right\rangle$ are given in (\ref{1}) in the points $\pm\Phi$ where the observers are placed. This final state of the correlated particles becomes analogue to the final state obtained in \cite{bf} when infinitesimal Lorentz boosts are applied at each point of the spacetime. We can see easily that the angle $\theta$ in the expression (\ref{4.8}) is identical to the angle of the Wigner rotation obtained in \cite{bf}, whose value is
\begin{eqnarray}
\theta=\pm\,i\gamma=\alpha\,\Phi\,\cosh\xi.
\label{4.9}
\end{eqnarray} 

At first sight, the final state (\ref{4.8}) indicates a deterioration of the perfect initial spin anticorrelations, if the observers make their spin measurements in the $3$-axis of its local reference frame. Actually, it does not indicate the deterioration of the initial spin anticorrelations (\ref{4.1}). The final state (\ref{4.8}) shows us that there is a precession about the $2$-axis in the local reference frame of observers, when they are placed at the points $\varphi=\pm \Phi$ of the spacetime, in the rest frame of the particles, as in \cite{bf}. This precession of spins, given by the phase shift of the Fermi-Walker transport, is analogue to the angle of the Wigner rotation on the spin axis of measurements, \textit{i.e.}, about the $2$-axis in the local reference frame of the observers. The angle of the Wigner rotation (\ref{4.9}) comes from the influence of the topology of the spacetime, the accelerated motion of the particles and the position of the observers. Hence, with the information about the topology of the spacetime and the motion of the particles, we always recuperate the perfect initial anticorrelation if we rotate the spin axis of measurements through the angles $\mp\theta$ about the $2$-axis in the local reference frames of the observers placed at the points $\varphi=\pm\Phi$.  

\section{bell inequality}
$ $ 

Let us analyze the violation of the Bell inequality in this system, where the particles are moving in a circular motion around the symmetry axis of the cosmic string. Suppose that the first observer placed in $\varphi=+\Phi$ chooses to measure the  spin component $\hat{a}$ or $\hat{a'}$ and the second observer placed in $\varphi=-\Phi$ chooses to measure the spin component $\hat{b}$ or $\hat{b'}$, where
\begin{eqnarray}
\hat{a}=\frac{\sigma^{1}+\sigma^{3}}{\sqrt{2}};\,\,\,\hat{a}'=\frac{-\sigma^{1}+\sigma^{3}}{\sqrt{2}};\,\,\,\hat{b}=\sigma^{3};\,\,\,\hat{b}'=\sigma^{1}.
\label{5.1}
\end{eqnarray} 

Taking the final state (\ref{4.8}), the violation of  Bell's inequality becomes
\begin{eqnarray}
\left|\left\langle ab\right\rangle+\left\langle a'b\right\rangle+\left\langle ab'\right\rangle-\left\langle a'b'\right\rangle\right|=\frac{2}{\sqrt{2}}\left|-\cos\left(2\theta\right)-\cos^{2}\theta+\frac{\left(\eta^{2}_{1}+\eta^{2}_{2}\right)}{\gamma^{2}}\,\sin^{2}\theta\right|
\label{5.2}
\end{eqnarray}

We can see that the degree of violation of the Bell's inequality decreases if the observers make their spin measurements on the $3$-axis in each local reference frame. The expression for the Bell inequality given by the application of the Fermi-Walker transport is different from the expression obtained by Terashima and Ueda \cite{TU2} and by the present authors in \cite{bf} under the application of successive infinitesimal Lorentz transformations. In all cases, the degree of violation of the Bell inequality obtained in \cite{bf} and by the Femi-Walker transport is caused by the accelerated motion of the particles, the geometry/topology of the spacetime, the position of the observers and the choice of the spin axis of measurements made by the observers.  However, we can restore the maximal violation of the Bell inequality in all cases if the observers rotate their spin axis of measurements with appropriated directions, that is, with an angle $\mp\theta$ around the $2$-axis in their local reference frame.    

\section{conclusions}
$ $
We study the relativistic EPR correlations in a curved spacetime background via application of the Fermi-Walker transport. We analyzed a \textit{gedankenexperiment} in the cosmic string spacetime where an EPR pair of particles was moving in circular motion around the symmetry axis of the cosmic string. In this way, we had assumed the WKB approximation and saw how the spatial part of the wave function of each particle behaves when it is transported via Fermi-Walker transport around the symmetry axis of the cosmic string. 

We conclude that the Fermi-Walker transport determines the rate at which the spatial part of the wave function is precessing relative to the local reference frame of the observers. The phase acquired by the wave function due the Fermi-Walker transport is identical to the angle of the Wigner rotation obtained by the application of infinitesimal Lorentz transformations \cite{bf} in the cosmic string spacetime. The influence of the topology of the spacetime, the motion of the particles and the position of the observers in the EPR correlations arises clearly with the Fermi-Walker transport as obtained by the authors in \cite{bf} through the application of successive infinitesimal Lorentz boosts. 

We saw that the violation of  Bell's inequality decreases with the application of the Fermi-Walker transport due the accelerated motion of the particles, the position of the observers and the topology of the spacetime, if the observers make their spin measurements on the $3$-axis in each local reference frame. However, the maximal violation of the Bell's inequality is restored if the observers rotate their spin axis of measurements about the angle $\mp\theta$ on the $2$-axis in their local reference frames.

We finish by commenting that a similar study of EPR correlations under the influence of torsion, which appears associated to line defects called dislocations\cite{37} is under  way.

\acknowledgments{ This work was partially supported by CNPq, CAPES (PROCAD), PRONEX/CNPq/FAPESQ and CNPq/Universal.}

\end{document}